# MODELING CONCORDANCES OF COMPANY'S INVESTMENT DIRECTIONS WITH ITS MARKET ATTRACTION

*V.Ya. Vilisov*
*University of Technology, Russia, Moscow Region, Korolev*
*vvib@yandex.ru*

**Abstract.** *This work models the interconnection of company's investment managers' representations and the market attraction of its shares. The models that reflect the connection of the company's market effectiveness indices and parameters of its economic activity are created on the basis of the Mean-Variance Analysis and Regression Analysis. On another side, expert evaluation methods also clarified the same influence parameters, but it was made according to the opinion of company managers. These two evaluation rows are used when making managerial decisions.*

### Introduction

Investment managers play a significant role in the work of listed companies, being the employees who choose the directions of the funds' investments in order to provide the competitiveness, capitalization and integral success of the company. In this situation it is quite important for them to be aware of the current business environment and strategic development trends of the company. They should possess error-free intuition, feeling the interconnection of the "control levers" for which they are responsible, which include not only the investments but also the external manifestations of the company's effectiveness, for example market value of its shares. Nowadays this kind of feedback exists only in the form of the managers' intuitive and highly uncontrolled representations, thus adding a lot of uncertainty and mumbo jumbo to their activity. The work tries to formalize the feedback, providing the company's top management with the tool to monitor the effectiveness of the internal investments.

### Object of Research

Joint Stock Companies (JSC) participate in two investment processes of the market economy. On one hand, shares of JSCs are quoted on the trading floors and the more effective and competitive the JSC is, the higher the market value of its shares is together with the competitiveness and capitalization. At that, the external investors tend to include the JSC shares into their investment portfolios when their profitability levels are high and risk levels are low. On another hand, internal investment funds are distributed among the different development directions pertaining to the company's activity. Should these distributions be effective, the external investors shall highly appraise the market attraction of JSC shares at the trading floors. Company's top management is responsible for the internal investments distribution. Thus, if its managerial decisions lead to the growth of the market attraction (as well as the capitalization) of the company, it means that it is adequate in feeling the connection of the managerial "levers" and the market reflection of the company's effectiveness. Should the management's actions result in the decrease of the company's market attraction, it is necessary to improve the quality of the management by making the staff shifts, increasing employees' professionalism, providing specific additional research etc.. Thus, the objects of the research are both the companies, whose shares are quoted on the trading floors and the company management that manages internal investments, making the company more market-attractive.

The research is performed on the basis of one machine-manufacturing JSC, hereinafter referred to as the Company. The Company's shares are quoted at the Russian Commodities and Raw Materials Exchange [8].

### Subject of the Research

This work tries to evaluate the perception adequacy of the company's (JSC) management in relation to the interconnection of internal factors and the company's market attraction. Currently this interconnection does not exist in the clear (formalized) form. On one hand, there is the market where external investors (in relation to the company) prioritize companies on the basis of their effectiveness (via the market value of their shares), i.e. the market evaluation of the company is an external, unprejudiced and a highly independent estimate. On another hand, company managers control internal levers (including the investments) by relying on their own experience and personal vision of the interconnection found between the internal factors and the company effectiveness, seen externally. At that their subjective opinion of effectiveness can be different from the unprejudiced market evaluations.

Should the incongruity be a significant one, it is clear that there exists a necessity to somehow correct the "internal models of managers" by sending them to a training, providing them with the additional information or by making changes within the managerial team. When managers inadequately perceive the interconnection of the managerial solutions for which they are responsible and the target function (for example, it could also be the integrated index of the company's market effectiveness), it means that they use corrupted managerial targets or they deliberately replace them with other targets [1]. New institutional theory explains the latter as the opportunistic behavior of managers [2].

Therefore, it is the hidden mechanism that provides unseen interconnection of the company's market effectiveness and its internal factors that can be perceived as the subject of the research (see Picture 1). Thus, in our opinion, the presence and/or necessity of such interconnection is highly apparent as practically every company that produces goods, intends to sell them at the corresponding market.

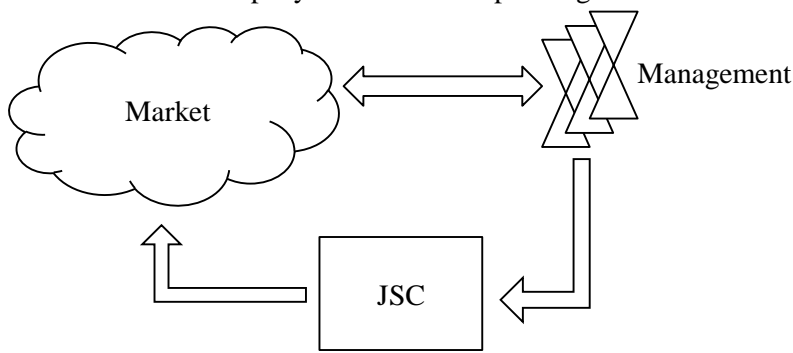

Pic. 1. Market Interconnection.

**Objective of the Research**

In order for the actions undertaken by the management in the course of the company management be effective, providing market attraction competitiveness and capitalization, it makes economic sense to possess the tools that could evaluate how accurately the managers understand both the management factors and their impact upon the company's effectiveness.

The objective of the work lies in the construction of the algorithm that could reveal the impact degree of some internal factors related to the company on the market data of the company shares.

**Source Data**

The source data for the analysis include: data collations of the Russian Commodities and Raw Materials Exchange, Company's quarterly reports that it openly publishes on the website as well as the data received from the expert questionnaires, filled in by its managers.

**Research Outline**

It is suggested that the impact degree of the factors should be evaluated with 2 methods (Pic. 2):

1. In order to draw unprejudiced evaluations: according to the market data of the Russian Commodities and Raw Materials Exchange and according to the company reports.

2. In order to draw subjective evaluations: according to the opinion of the company management.

The evaluation results that are obtained by these two methods shall be compared so that top managers could make managerial decisions in relation to the staff, authorities and further work of the line manager team using the comparison data. The work contains economic and mathematical means of modeling the mentioned elements basing on the available data.

Further we shall give consideration to both evaluation methods, where, within the framework of each one of them we shall review their mathematical problem definitions, descriptions of the solution algorithms, procedures used for the receipt of the interim and final evaluations made on the basis of the actual statistical data, also providing brief comments regarding the obtained results.

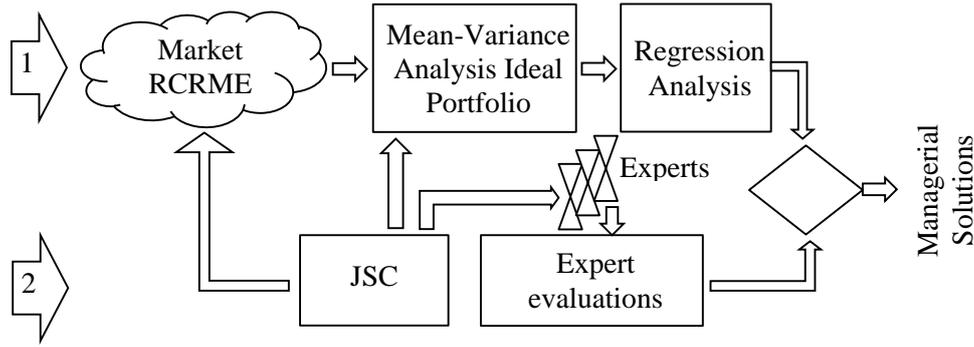

Pic. 2 Modeling Pattern

**Mathematical Models of Unprejudiced Evaluation (First Method)**

Though this method is called unprejudiced, it is based on the actions of the persons who make decisions regarding buying the shares at the trading floors. However, taking into consideration the fact that market entities do not depend on the company's managers, being the elements of the external and independent environment in relation to the Company, they do reflect an unprejudiced surrounding of the company.

The first method is based on the 2-type models:

1. Mean-Variance Analysis Model [3-7], which provides a possibility to evaluate the Company Share Fraction (CSF) contained within the investment portfolio of some sensible external investor.

2. Regression Model that connects CSF and some internal factors of the Company, whose values are provided in the regularly published Company reports and which can be influenced by the managers that control some activities or development of this Company, in particular dealing with the investment distribution.

**Mean-Variance Analysis Model**

The value of the Company Share Fraction contained within the ideal portfolio (IP) according to the Mean-Variance Analysis was chosen as a unital integrated index of the Company's market attraction.

Let us briefly consider the main components of the IP. Let us assume that the external investor defined the range of the securities that are potentially suitable for their inclusion into the portfolio. In this case the problem of IP formation is to make such a security portfolio (choose the $\bar{x} = [x_1 \ x_2 \ \cdots \ x_n]^T$ vector) that would provide the minor risks $\sigma_p$ (or variance $D_p = \sigma_p^2$) with the set profitability being $m_p$. Here $x_i, i = \overline{1,n}$ is the share of investments contained within the securities of $i$–type, with $T$ being the conjugation symbol.

The optimization criterion applied to the IP search problem takes the following form:

$$D_p = \sum_{i=1}^{n}\sum_{j=1}^{n} x_i x_j K_{ij} \to \min_{\bar{x}}, \qquad (1)$$

where $K_{ij}$ is a covariance of two kinds of securities: $i$-type and $j$-type.

But delimitation serves as the condition of congruence of the expected profitability to some desired level $m_p$:

$$\sum_{j=1}^{n} x_j m_j = m_p, \qquad (2)$$

where $m_j$ is the average (expected) profitability for the securities of $j$-type. All desired investment fractions should meet the requirements of the normalization condition:

$$\sum_{j=1}^{n} x_j = 1. \qquad (3)$$

When in the vector-matrix form, the problem (1)-(3) takes the form of:

$$D_p = \bar{x}^T K \bar{x} \to \min_{\bar{x}}, \qquad (4)$$

with the delimitations being:

$$\bar{m}^T \bar{x} = m_p. \qquad (5)$$
$$I^T \bar{x} = 1. \qquad (6)$$

Here:
$K = \|K_{ij}\|_{nn}$ - covariance square matrix of the reviewed set of securities;

$\bar{m} = [m_1 \ m_2 \ \cdots \ m_n]^T$ - vector of expected (average) profitabilities of the securities set;

$I = [1 \ 1 \ \cdots \ 1]^T$ - unital vector.

This problem was solved on the basis of the source data, described below.

**Source Data Comments**

The research is reviewing the period from the 4th quarter of 2006 until the 4th quarter of 2011 as all quarterly financial reports are available to the public (on the Company's website). Thus, we have reviewed 21 quarter-based data set. Share quotation source data was taken from the website of the Russian Commodities and Raw Materials Exchange [8].

The JSCs, whose shares were considered for inclusion into the Mean-Variance Analysis, are the following:
1. Company;
2. Gazprom;
3. Lukoil;
4. Sberbank;
5. Rostelecom;
6. Rosneft;
7. Uralkali;
8. Norilsk Nickel;
9. Aeroflot;
10. Severstal.

Further, for the purposes of briefness, we shall use numbers instead of the company names. Statistic data on profitability (expressed as the percentage off the nominal cost of the company shares) is provided on Pictures 3-5. Securities classification diagrams related to their profitability and risks are given as average values for the research period under review.

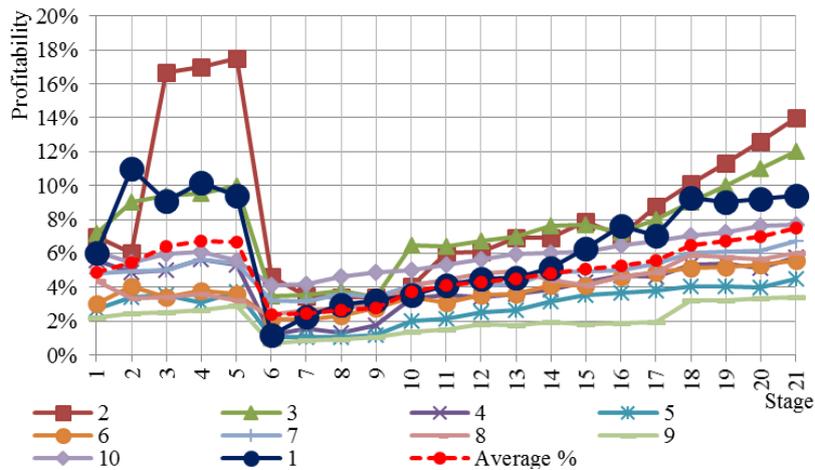

Pic. 3. Quarterly Profitability Dynamics of Company Shares

Points related to the research stages, which are shown on Picture 3, express average profitabilities of the securities. They were calculated on the basis of all trade sessions performed during the research period under review (of the current quarter). The points from No. 5 to No. 9 refer to the period of the financial crisis of 2008. Point No. 10 is the 1st quarter of 2009, starting from which it is possible to notice stable and remedial profitability growth for all shares of the reviewed group.

Given that the research was performed during the highly variable period (crisis and recovery), it is evident that the statistic characteristics that are necessary for the Mean-Variance Analysis (average profitabilities and covariances of the securities) are also rather variable. Considering this, the work contains the means that allow to balance the impact of the variability factors.

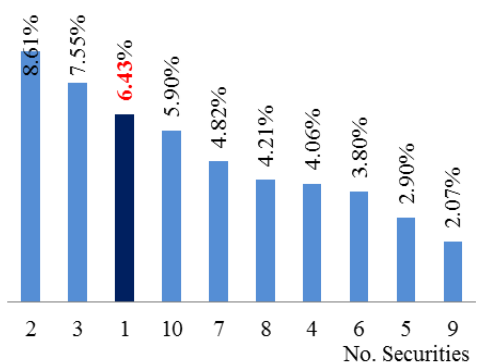

Pic. 4. Securities Profitability

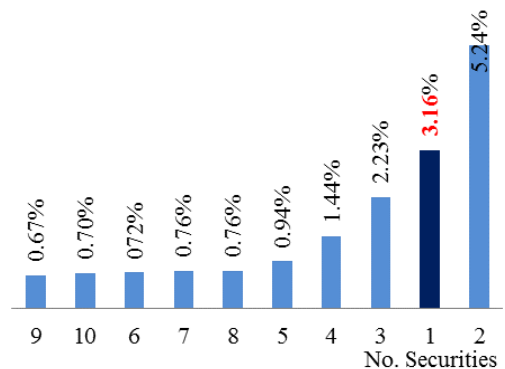

Pic. 5. Risks of Securities

First of all, first 9 points possessing the most variability were excluded from the review, even though thus we have considerably reduced the sample volume. Secondly, when computing the covariance matrix $K$ that is necessary for the calculations, we have excluded the linear component (trend) for the rest 12 points, thus increasing the calculation accuracy of the paired covariances.

Values of Company indices are expressed in dark colour (see Pictures 4 and 5) on the classification diagrams.

As the external investor prefers to have high-profitability and minimal risk securities in his portfolio, the securities of the Company (including other companies of the pool) cannot be considered ideal, because together with the high profitability they possess major risks.

**Computing IP Parameters**

The IP building problem was solved with the consideration of the above-mentioned transformations of the statistic data for the last 12 research points. As is known, the IP building problem lies in the provision of the desired degree of the portfolio profitability $m_p$ with the minimized risks. Given that the desired profitability can lie within the interval ranging from the lowest (out of the whole securities pool under review) till the highest, it would be only natural to suggest that the degree of profitability, acceptable for some investor, does not lie within the extreme points of this interval.

It would be possible to define the desired level of profitability by solving the portfolio stability maximization problem using the research interval, but we shall not consider this management aspect in this work.

Let us agree that when investing the funds into the portfolio, the investor expects to receive above-the-average profitability. For the purposes of determination in this work, it is accepted that during the whole research period, the desired level of profitability, received by the external investor, constitutes $m_p = 0.75$ starting from the lowest border of the profitability interval referring to the current moment of the research.

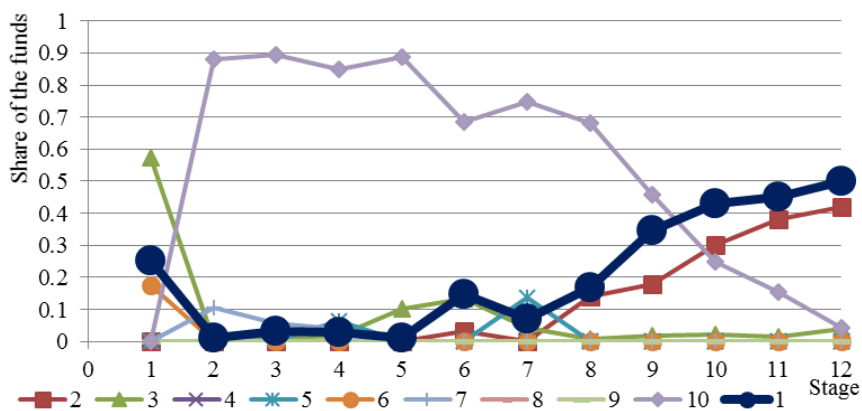

Pic. 6. Fractions of Company Shares in the Mean-Variance Analysis Portfolio.

After taking into consideration the above-mentioned comments, the IP selection problem was solved in MS Excel environment using the Solver tool for each of the 12 mentioned research stages. The above-mentioned expressions for the target function and delimitations (1)-(3) were used in the computations. The received fractions of the securities taken from the pool of 10 companies under the review are provided on Picture 6 with the modification trajectory expressed in bold. The trajectory shows the quarterly portfolio modifications of the partial funds, invested into the Company's securities. The picture shows that the portfolio's Company share started growing only during the period of 2010-2011.

**Some Comments on the Securities Pool Composition in the Portfolio**

In the real time, multitude of the securities that are reviewed within the composition of the portfolio's pool, can change depending on the moment of the research. There can be many companies, whose securities could be included into the pool, but it does not make economic sense to have many securities in the portfolio, as it will hinder its management. There should only be a limited number of securities left within the pool composition. At that the trader should be able to reason why some securities are excluded from the portfolio while others are included. In our opinion, such reasoning can be made with at least one of the following two methods:

1. To include all available securities and to solve the IP problem, afterwards excluding the securities, whose share was close to zero. Then the IP problem should be re-solved using the reduced pool. At each research stage there should be several screening iterations.

2. When reviewing maximum composition of the initial securities pool, it is possible to use the convexity property of the portfolio multitude within the profitability-risk coordinates. At that, with the consideration of the current (at the corresponding research stage) values of the profitability and risks, it is advised to emphasize the Pareto line [9, 10], where its security components shall constitute the current pool, basing on which the IP problem should be solved.

The second variant of the securities pool management is considered to be more absolute and faithful. Such pool, built with the securities of the Pareto line and applied for the data under the review is provided on Picture 7. However, it is necessary to note that in this case the Company's securities (point 1) are not located on the Pareto line, but there are no contraindications for them not to be included in the pool together with the securities No. 2, 3, 9 and 10.

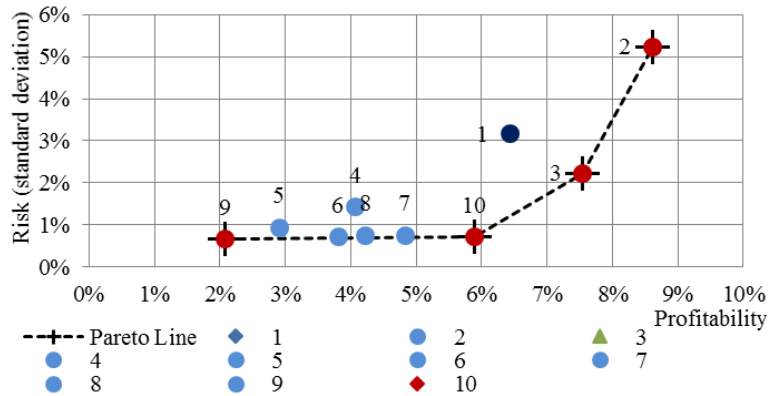

Pic. 7. Pareto Line in the Portfolio Multitude

Still, in this work we shall not consider issues of the securities pool management at each stage of the research, supposing instead, that the securities pool is constant.

**Building Regression Model**

This stage of the analysis is the second part of the first method (unprejudiced evaluation) that investigates the connection of the outgoing indices and internal Company factors.

The idea of this analysis consists in the following: to calculate, using the unprejudiced data, the degree of impact of some internal Company factors upon its shares, included into the IP at each stage of the research.

These factors should be open for investigation and be available for calculation and registration at each stage of the research. The source data for this work consisted of the quarterly reports, which are published on the Company's official website and which contain all main parts of the balance sheet (income and expense statement, cash flow statement etc.). They also contain information regarding the investments into the different spheres of the Company's activity and development. Should it be necessary, this data package could be investigated in detail with the consideration of different parameters, but in this work, for the purposes of keeping the generality, we shall consider only five parameters. In our opinion, it is these parameters that are influencing the Company's market indices, including the index under review, i.e. fraction of the Company's securities in the IP. The parameters under review are the following:

1. Fixed assets total ($f_1$).
2. Gross payroll ($f_2$).
3. Net income total ($f_3$).
4. Profit margin ($f_4$).
5. Major produce throughput rate ($f_5$).

Given that all factors have different meanings and are measured with different units, let us use a factor normalization method in order to consubstantiate them. Normalization is performed in such a manner that each of the statistic data variables is defined with the minimum ($f_i^{min}$) and maximum ($f_i^{max}$) values, determining the borders of the variability interval per each variable (factor). Within the reference scales each interval is then assigned with the non-dimensional interval [0;1], thus each value of any factor under review shall correspond to the normalized value located within the range. Table of the factors' normalized values referring to all 12 stages of the research as well as the values of the Company shares ($x_1$) within the IP are provided in Table 1.

Table 1. **Normalized Values of the Factors on All Stages of the Research**

| Stage | Share ($x_1$) | $f_1$ | $f_2$ | $f_3$ | $f_4$ | $f_5$ |
|---|---|---|---|---|---|---|
| 1 | 0.251 | 0.000 | 0.000 | 0.317 | 0.714 | 0.667 |
| 2 | 0.012 | 0.007 | 0.000 | 0.000 | 0.571 | 0.333 |
| 3 | 0.033 | 0.086 | 0.000 | 0.124 | 0.429 | 0.333 |
| 4 | 0.029 | 0.287 | 0.000 | 0.286 | 0.286 | 0.667 |
| 5 | 0.012 | 0.247 | 0.000 | 0.617 | 0.143 | 0.000 |
| 6 | 0.148 | 0.315 | 0.224 | 0.122 | 1.000 | 1.000 |
| 7 | 0.070 | 0.485 | 0.224 | 0.335 | 0.143 | 0.333 |
| 8 | 0.170 | 0.757 | 0.224 | 0.538 | 0.000 | 0.333 |
| 9 | 0.347 | 0.698 | 0.224 | 0.887 | 0.571 | 0.333 |
| 10 | 0.428 | 0.759 | 1.000 | 1.000 | 0.857 | 0.333 |
| 11 | 0.452 | 0.830 | 1.000 | 0.307 | 0.143 | 0.667 |
| 12 | 0.501 | 1.000 | 1.000 | 0.999 | 0.286 | 0.667 |

The research data was processed using the Regression option of Data Analysis, the MS Excel add-in. As a result of this, we received the following regression equation:

$$x_1 = c_0 + c_1 f_1 + c_2 f_2 + c_3 f_3 + c_4 f_4 + c_5 f_5$$
$$= -0.075 - 0.006 f_1 + 0.262 f_2 + 0.216 f_3 + 0.029 f_4 + 0.179 f_5. \quad (7)$$

It is necessary to note that due to the fact that determination coefficient is $R^2 = 0.87$, the regression model has a rather high adequacy level. However, coefficients at $f_1$ and $f_4$ are not significant. As is known, regression equation coefficients reflect both the degree of the factor-output value correlation relationship and each factor's contribution into the total effect (Company's share fraction in the IP). The factors, prioritized according to their impact degree upon the output value, are provided on Picture 8.

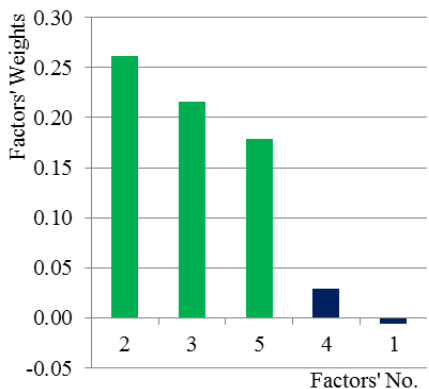

Pic. 8. Factors' Weights According to the Regression Model

Thus, the result of the unprejudiced evaluation is the weights of the internal factors, received with the regression analysis. Therefore, the first method out of the two, provided on Picture 1, was already implemented. Let us implement the second method.

**Subjective Evaluation of the Company's Internal Factors Contribution into the Integral Effect of its Work (Second Method)**

This evaluation method is based on the opinion of the Company's managers. Usually, it is considered that while distributing investment funds and performing other managerial actions, Company management can adequately foresee the impact of some managerial "levers" onto the final effect. First of all, it is based on their experience and managerial professionalism. Managers should feel the connection intuitively, but in real life it does not always happen so ideally. This is why the Company top managers should be aware of how adequately the line managers understand the impact degree of some managerial levers onto the final effect. This is where the determination of the Company managers' weight impact coefficients plays the important role, providing a further comparison of the obtained data and the weight evaluations, which were received with the first method, i.e. according to the "market opinion". If the incongruity is significant, it will serve as a signal to the Company top managers to take some organizational measures for setting up the management towards the market's opinion.

In order to build subjective evaluations of the internal factors' contribution we used an expert evaluation technique [11], according to which:

1. We have formed an expert group consisting of the Company's line managers, whose functional responsibilities included activities on investment management, oriented at the growth of the Company's market capitalization. The group consists of 10 managers.

2. We have developed a questionnaire, which, on the basis of the answers obtained from the experts, allows building paired comparison matrices defining the degrees of importance (significance) of the internal factors' influence, which are mentioned above, upon the market value of the Company shares. The experts provided their answers in the two scales: in the discrete scale (Yes/No) and in the continuous scale, where they gave a percentage value when defining the significance degree of the factor pairs. Two-variant data received from the experts allows to reduce the subjective evaluation errors.

3. Every paired comparison matrix (PCM) is processed with the methods of Lewis, Summation and Multiplication, thus allowing to reduce the mathematical errors when evaluating the factors' weights. Using the values of the experts' number and processing methods, we have calculated average weight coefficients that reflect each factor's significance or impact degree upon the market quotes of the Company shares.

The binary scale questionnaires contained the following question to be answered by the experts: "When quoting the Company shares at the Russian Commodities and Raw Materials Exchange, will the first factor out of the two provided be more significant than the second? (Yes/No)". At that the expert should have put some sign (+, *, ...) in the column "Yes" or "No". All questions and filled in variants are provided in Table 2.

Table 2. **Binary Questionnaire Filling In Example**

| PCM Indices | No. | Yes | No | More significant factor | Less significant factor |
|---|---|---|---|---|---|
| 1;2 | 1 | + | | Fixed assets total | Gross payroll |
| 1;3 | 2 | + | | Fixed assets total | Net income total |
| 1;4 | 3 | + | | Fixed assets total | Profit margin |
| 1;5 | 4 | + | | Fixed assets total | Major produce throughput rate |
| 2;3 | 5 | | + | Gross payroll | Net income total |
| 2;4 | 6 | | + | Gross payroll | Profit margin |
| 2;5 | 7 | + | | Gross payroll | Major produce throughput rate |
| 3;4 | 8 | | + | Net income total | Profit margin |
| 3;5 | 9 | + | | Net income total | Major produce throughput rate |
| 4;5 | 10 | + | | Profit margin | Major produce throughput rate |

It is necessary to mention that the questionnaire did not include Table 1's first column. Here it is added in order to show the connection of the question's number and the PCM cell, thus the first index stands for the PMC line number and the second for the column number. When filling in the PCM, "Yes" value corresponds to value "2" and "No" to "0". Should the values be equivalent, the value is "1". When answering the questionnaire, the upper triangle of PCM is to be filled (Table 3, highlighted gray), with the lower filled as an addition to the upper. For example, the value "2" in the cell (1;2) of the lower triangle should correspond to the value "0" of the cell (2;1) of the lower triangle and so forth.

According to the questionnaire data we built PCM, one of which (built according to Table 2) looks like the one in Table 3. Table 4 contains PCM, built according to the questionnaires, which were filled in by the same experts but using a continuous scale (percentage is worked out to [0;1] interval).

Table 3. **Discrete PCM**

| Factors | Factors | | | | |
|---|---|---|---|---|---|
| | 1 | 2 | 3 | 4 | 5 |
| 1 | 1 | **2** | **2** | **2** | **2** |
| 2 | 0 | 1 | **0** | **0** | **2** |
| 3 | 0 | 2 | 1 | **0** | **2** |
| 4 | 0 | 2 | 2 | 1 | **2** |
| 5 | 0 | 0 | 0 | 0 | 1 |

Table 4. **Continuous PCM**

| Factors | Factors | | | | |
|---|---|---|---|---|---|
| | 1 | 2 | 3 | 4 | 5 |
| 1 | 0.5 | **0.85** | **0.3** | **0.4** | **0.2** |
| 2 | 0.15 | 0.5 | **0.1** | **0.2** | **0.1** |
| 3 | 0.7 | 0.9 | 0.5 | **0.6** | **0.55** |
| 4 | 0.6 | 0.8 | 0.4 | 0.5 | **0.3** |
| 5 | 0.8 | 0.9 | 0.45 | 0.7 | 0.5 |

The questionnaire question, waiting for the continuous scale answer, is the following: "When quoting Company shares at the Russian

Commodities and Raw Materials Exchange, how would you distribute 100% of significance (impact upon the shares' cost) between factor pairs (% of the 1st factor + % of the 2nd factor = 100%)".

Continuous scale table, provided for the experts to fill it in, was similar to Table 2. The only difference is that the experts should have inserted corresponding percentage instead of "Yes" and "No" answers. Continuous scale PCM are similar to the ones in Table 4.

For the purposes of processing discrete PCM, we only used the Summation method as the discrete scale is too rough for other methods. Continuous scale PCM were processed with all three methods (Summation, Multiplication, Lewis method). Thus, each expert provided (after processing) 4-method factor weight vector: discrete scale summation, continuous scale summation, continuous scale multiplication and by continuous scale using Lewis method. We also checked transitivity of the obtained evaluations.

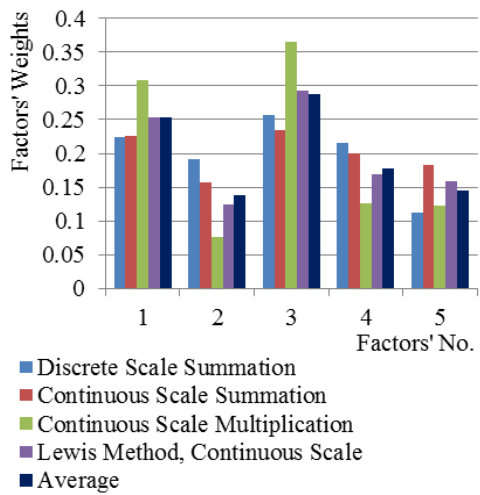

Pic. 9. Average Weights of Factors.

Using all these versions of weight evaluations, each expert was then provided with the average value of these 4 versions' multitude, which further were averaged out according to the experts' multitude. Picture 9 contains weights of factors, which were averaged out according to the experts' multitude. Picture 10 contains values, which were ranked according to the weight decrease and averaged out according to the multitude of the processing methods.

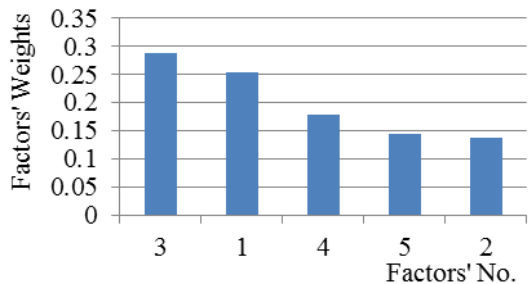

Pic. 10. Factors' Expert Evaluations.

The reliability issue is very important for the obtained evaluations. Given that every evaluation was calculated as an average value in relation to its processing methods and number of experts (when calculating each evaluation, the sample consisted of 40 observations), the significance check showed that all weight evaluations are significant.

**Comparison of Unprejudiced and Subjective Evaluations of the Factors' Significance**

Weight evaluation values and ranking, calculated according to the unprejudiced data (see Pic. 8) differ from the data that was obtained as a result of the expert poll (see Pic. 10). In order to perform a more accurate comparison of the significance of the Company's internal factors, let us present them both in Table 5 and the integrated diagram (Pic. 11). Here we should mention that the weights, obtained with the 1st method are standardized in such a manner that their sum would be equal to 1, therefore their values are slightly different from the ones that were obtained initially (see Equation (7) and Pic. 8). Weight evaluations that were obtained with the 2nd method, were standardized initially.

Table 5. **Weight Evaluations of Unprejudiced and Subjective Factors**

| Evaluation Methods | Factors | | | | |
|---|---|---|---|---|---|
| | 1 | 2 | 3 | 4 | 5 |
| Unprejudiced (Russian Commodities and Raw Materials Exchange) | -0.008 | 0.384 | 0.317 | 0.043 | 0.263 |
| Subjective (experts) | 0.253 | 0.138 | 0.287 | 0.178 | 0.144 |

It is necessary to mention that the evaluations of the factors' significance obtained from the multitude of the real data possess a rather high reliability level. Factor significance evaluations that were obtained with these two methods are completely different.

For the sake of convenience of two groups of evaluations it is desirable to have some scalar disparity measure. A correlation coefficient value found between two factor significance evaluation groups could be here a standard measure, where should they completely coincide, the value is $r_f = 1$, should the noncorrelatedness be complete, the value is $r_f = 0$ and should the evaluations be opposite, the value is $r_f = -1$. Let us consider this weight variant using standard functions of MS Excel. As a result of the calculations performed for two samples, provided in Table 5, the correlation coefficient is $r_f = -0.27$, the value of $t$-statistics is $t = 0.48$, where significance probability of the obtained correlation coefficient is only $P = 0.34$. It is obvious that this low level of the probability belief is caused by the very small sample (only five factors). Thus, correlation coefficient can serve as a measure of the evaluations' concordance or discordance. It can be calculated for a different number of factors, but its belief degree (probability belief) is still low in relation to the small samples.

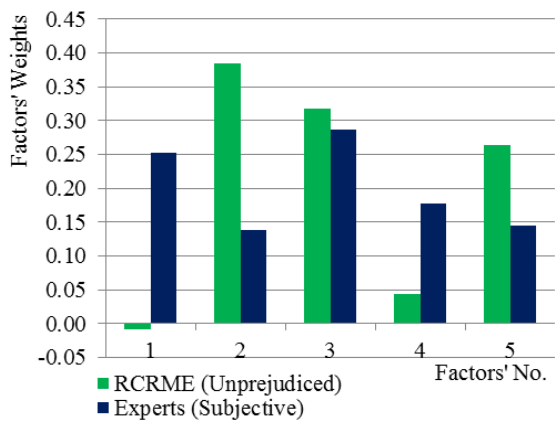

Pic. 11. Unprejudiced and Subjective Factor Weights

Measures that reflect the deviation degree of two evaluation types can include variability $D_f$ or root-mean-square deviation (RMSD) $\sigma_f$ of the difference between the evaluations of the factors obtained with two different methods. At that, should the evaluations of two types coincide, it is $\sigma_f = 0$. Maximum value of RMSD is limited by 1: $\sigma_f = 1$. Value of this weight variant, calculated with the use of MS Excel standard functions, constitutes $\sigma_f = 0.201$, which is a rather significant value, because average values of the factor evaluations obtained by two different methods are the same and constitute $m_f = 0.2$.

Therefore, the two methods (correlation and variable) used for comparison of two evaluation groups of the significance coefficients (subjective and unprejudiced) do not permit to consider them as close enough.

**Conclusions**

1. The difference between the subjective and unprejudiced factor significance evaluations is indicative of the perception inadequacy that exists among the Company management regarding the market mechanisms, also reflecting the fact of the non-effective investment company management.

2. Significant deviations between two evaluation groups can serve as a basis for correcting the financial policy of the company or for replacing the line manager staff.

3. This analysis method can be used as tool to employ the managers who possess an adequate understanding of the Company's working mechanisms within the market environment.

4. The analysis method proved its work capacity, the potential to use in practice with a more expanded number of factors as well as the potential to include a larger number of experts.

**Discussion**

Let us define the following further directions of the research:

1. As a rule, investment targets possess a more complicated structure in real life, including much more than just an increase of the company capitalization. In the majority of cases, there exists (clear or hidden) a hierarchy of targets, which, apart from purely financial, can also include technical targets (which, quite likely, indirectly influence the capitalization) containing of the manufacturing technology update, familiarization with new samples etc. Discovery of patterns related to the structural and temporal interconnection between the separate effects of the investments and the total integral indices presents a practical value, but can be accomplished only with the help of the concrete statistical material of an existing JSC. Apart from that, it is necessary to consider the fact that different companies can have a different impact upon the final indices of the capitalization: ones have a higher impact coefficient, others have a lesser impact coefficient, ones become apparent quickly while others become apparent later. Therefore, in order to perform a more complete research of the mechanisms that reflect an interconnection of the internal investment processes and the market (external) attraction of the Company shares, it is necessary to make an analysis of the managerial target hierarchy and to discover the groups of factors that influence the market indices of the Company, also discovering the dynamic properties of the different factor groups (lateness, aperiodicity, variability etc.).

2. Given that the main idea of the explained approach is based, for the purposes of the managerial decision-making, on the ultimate feedback i.e. the final integral effect (the unprejudiced effect, which was calculated in the normal operation mode within the conditions of the mature market), it is also important to research other variants of such external demonstration of the integral effect (apart from open trading floors like the Russian Commodities and Raw Materials Exchange). Such analysis is important for the companies that do not possess the open trading floors access capability (small and middle companies, defense enterprises, science organizations etc.).

3. Construction of the system involving mathematical models, program means, information resources as well as the organizational and managerial procedures united into one human & machine complex, responsible for monitoring the effectiveness of the investment processes and supporting managerial decision-making, including the preparation of the decision variants, is also deemed very important.

4. Existence of the mathematical model, reflecting the interconnection of the market demand integrated index of the Company shares (for example, their fraction within the investment portfolio) and the internal Company factors allows building pure procedures of optimal distribution of the limited number of investments between the different activities (and/or development) based, for example, on the mathematical programming means or other pure methods of optimization.

**References**
[1] S. Beer. Brain of the Firm. - New York: John Wiley, 1988.
[2] O.E. Williamson. The Economic Institutions of Capitalism: Firms, Markets, Regional Contracting, 1985, Free Press, NY, 445 p.
[3] H.M. Markowits. Portfolio Selection, Journal of Finance. 1952. 7. № 1 pp. 71-91.
[4] H.M. Markowits. Mean Variance Analysis in Portfolio Choice and Capital Markets. Basil. Blackwell. 1990.
[5] G.J. Alexander. From Markowitz to modern risk management. European Journal of Finance, Vol. 15, Nos. 5–6, July–September 2009, pp. 451–461.
[6] S. Greco, B. Matarazzo, R. Słowin´ski. Beyond Markowitz with multiple criteria decision aiding, J Bus Econ, Feb. 2013, DOI: 10.1007/s11573-012-0644-2.
[7] V.A. Babaitsev, A.V. Brailov, V.Yu. Popov, Fast Algorithm for the Markowitz Critical Line Method, Mathematical Models and Computer Simulations, 2012, Vol. 4, No. 2, pp. 239–248.
[8] Russian Commodities and Raw Materials Exchange (Moscow Exchange), URL: http://moex.com/ (visited on 12.09.2015).
[9] H. Raiffa, R. Schlaifer. Applied Statistical Decision Theory. 2000.
[10] M.H. DeGroot Optimal statistical decisions. MeGraw Hill. N-Y. 1970. 253 p.
[11] T.L. Saaty, K.P. Kearns, E. Y. Rodin. Analytical Planning. The Organization of Systems, Pergamon Press, 1985, 210 p.